\documentclass[10pt]{article}

\usepackage{amsmath}

\usepackage{array}

\usepackage{appendix}

\usepackage{tocloft}                   

\usepackage{graphicx}

\usepackage{amsfonts}

\usepackage{amssymb}

\usepackage{mathrsfs}

\usepackage{yfonts}

\usepackage{euscript}

\usepackage{centernot}                 

\usepackage{ifsym}                     

\usepackage{lmodern}                   

\usepackage{upgreek}

\usepackage{mathtools}

\usepackage{color}

\usepackage{slantsc}
\usepackage{calligra}

\usepackage{bbold}          

\usepackage[T1]{fontenc}

\usepackage{epsf}

\usepackage{latexsym}

\usepackage{tipa}

\usepackage{makeidx}

\makeindex

\def\b{\begin{equation}}

\def\e{\begin{equation}}

\def\be{\begin{equation}}              

\def\ee{\end{equation}}

\def\beq{\begin{equation}}

\def\eeq{\end{equation}}

\def\bea{\begin{eqnarray}}

\def\eea{\end{eqnarray}}

\def\half{\mbox{$\frac{1}{2}$}}

\def\m{\mbox{ }}

\def\!{\hspace{-1.6667em}}

\def\c{\cite}

\def\n{\noindent}

\def\u{\underline}

\def\slLambda{\mathit{\Lambda}}                   

\def\biB{\mbox{\boldmath$B$}}              

\def\biG{\mbox{\boldmath $G$}} 

\def\biH{\mbox{\boldmath$H$}}

\def\biQ{\mbox{\boldmath$Q$}}

\def\bupxi{\mbox{\boldmath$\xi$}}                       

\def\bupchi{\mbox{\boldmath$\chi$}}                     

\def\sbupxi{\mbox{\scriptsize\boldmath$\upxi$}}

\def\mB{\mbox{B}}  

\def\mH{\mbox{H}} 

\def\mN{\mbox{N}}

\def\mZ{\mbox{Z}}

\def\me{\mbox{e}}

\def\mh{\mbox{h}}

\def\mp{\mbox{p}}

\def\bh{\u{\u{\mbox{h}}}  }            

\def\bN{\mbox{\bf N}}

\def\bh{\mbox{\bf h}}

\def\bp{\mbox{\bf p}}

\def\bcalD{\mbox{\boldmath ${\cal D}$}}

\def\sa{\mbox{\scriptsize a}}

\def\sf{\mbox{\scriptsize f}}

\def\si{\mbox{\scriptsize i}}

\def\sll{\mbox{\scriptsize l}}  

\def\sn{\mbox{\scriptsize n}}

\def\sr{\mbox{\scriptsize r}}

\def\st{\mbox{\scriptsize t}}

\def\su{\mbox{\scriptsize u}}

\def\sbN{\mbox{{\bf \scriptsize N}}}

\def\sbcM{\mbox{\boldmath \scriptsize ${\cal M}$}}

\def\bscM{\mbox{{\bf \scriptsize${\cal M}$}}}

\def\sumi2{\sum\mbox{}_{\mbox{}_{\mbox{\scriptsize $i$=1}}}^2}

\def\sumi3{\sum\mbox{}_{\mbox{}_{\mbox{\scriptsize $i$=1}}}^3}

\def\sumABcycles3{\sum\mbox{}_{\mbox{}_{\mbox{\scriptsize cycles $A,B$=1}}}^{3}}

\def\sumCDcycles3{\sum\mbox{}_{\mbox{}_{\mbox{\scriptsize cycles $C,D$=1}}}^{3}}

\def\sumj3{\sum\mbox{}_{\mbox{}_{\mbox{\scriptsize $j$=1}}}^3}

\def\sumk3{\sum\mbox{}_{\mbox{}_{\mbox{\scriptsize $k$=1}}}^3}

\def\prodiA1{\prod\mbox{}_{\mbox{}_{\mbox{\scriptsize $i$=1}}}^{A - 1}}

\def\pa{\partial}                                                   

\def\bpa{\mbox{\boldmath$\partial$}}                                                   

\def\es{\m = \m}

\def\:={\m := \m}

\def\=:{\m =: \m}

\def\FrM{\mbox{$\mathfrak{M}$}}                                

\def\sFrf{\mbox{\large $\mathfrak{f}$}}                          

\def\Hilb{\mbox{{\boldmath$\mathfrak{H}$}ilb}}                 
\def\FrL{\mbox{\boldmath$\mathfrak{L}$}}                       

\def\scH{\mbox{\scriptsize ${\cal H}$}}                    

\def\scM{\mbox{\scriptsize ${\cal M}$}}                    

\def\FrQ{\mbox{\Large $\mathfrak{q}$}}                               
\def\Phase{\mbox{{\boldmath$\mathfrak{P}$}hase}}                     

\def\bFrR{\mbox{\boldmath$\mathfrak{R}$}}                            
\def\Rig-Phase{\bFrR\mbox{ig-}\Phase}                                
                                                                													   
\def\bFrR{\mbox{\boldmath$\mathfrak{R}$}}                            

\def\bFrR{\mbox{\boldmath$\mathfrak{R}$}}                            

\def\1mat{\u{\u{1}}}                                                 

\def\Positive-Modespace{\mbox{{\boldmath$\mathfrak{M}$}odespace$^+$}}

\def\POSITIVE-MODESPACE{\mbox{{\boldmath$\mathfrak{M}$}ODESPACE$^+$}}

\def\Kin-Hilb{\mbox{{\boldmath$\mathfrak{K}$}in-\Hilb}}                     

\def\Mid-Hilb{\mbox{{\boldmath$\mathfrak{M}$}id-\Hilb}}                     

\def\Dyn-Hilb{\mbox{{\boldmath$\mathfrak{D}$}yn-\Hilb}}                     

\def\5Star{\mbox{\Large$\star$}}              

\begin{document}

\begin{center}

\Large{\bf Problem of Time and Background Independence:}

\vspace{.1in}

\normalsize

\Large{\bf classical version's higher Lie Theory}

\vspace{.1in}

{\large \bf Edward Anderson}$^1$ 

\vspace{.1in} 

\end{center}

\begin{abstract}

A local resolution of the Problem of Time has recently been given, alongside reformulation as a local theory of Background Independence.   
The classical part of this requires just Lie's Mathematics, much of which is basic:
i) Lie derivatives to encode Relationalism.
ii) Lie brackets for Closure 
   giving Lie algebraic structures. 
iii) Observables defined by a Lie brackets relation, 
   in the constrained canonical case as explicit PDEs to be solved using Lie's flow method, 
   and themselves forming Lie algebras. 
iv) Lattices of constraint algebraic substructures induce dual lattices of observables subalgebras.

The current Article focuses on two pieces of `higher Lie Theory' that are also required. 
Preliminarily, we extend Dirac's Algorithm for Constraint Closure to `Lie's Algorithm' for Generator Closure.
1) We then reinterpret `passing families of theories through the Dirac Algorithm' 
-- a method used for Spacetime Construction (from space) and getting more structure from less structure assumed more generally --
as the Dirac Rigidity subcase of Lie Rigidity. 
We also provide a Foundations of Geometry example of specifically Lie rather than Dirac Rigidity, 
to illustrate merit in extending from Dirac to Lie Algorithms.  
We point to such rigidity providing a partial cohomological (and thus global) selection principle for the Comparative Theory of Background Independence.
2) We finally pose the universal (theory-independent) analogue of GR's Refoliation Invariance for the general Lie Theory: 
Reallocation of Intermediary-Object Invariance.  
This is a commuting pentagon criterion: in evolving from an initial object to a final object, 
does switching which intermediary object one proceeds via amount to at most a difference by an automorphism of the final object?
We argue for this to also be a selection principle in the Comparative Theory of Background Independence.

\end{abstract}

$^1$ dr.e.anderson.maths.phyics *at* protonmail.com

\section{Introduction}\label{Introduction}

A local resolution \cite{ALett, ABook, I, II, III, IV, V, VI, VII, VIII, IX, X, XI, XII, XIII, XIV} 
of the Problem of Time \cite{Battelle, DeWitt67, Dirac49, Dirac51, Dirac58, Dirac, K92, I93, APoT, APoT2, ABook} has recently been given.
This refers to the eleven-facet version of the problem \cite{I, II, III, IV}: 
a slight repackaging of the nine-facet version \cite{APoT3, ABook} 
which extends the eight-facet version of Kucha\v{r} \cite{K92} and Isham \cite{I93} to allow for spacetime-centred as well as canonical approaches.  
Being `local', the global facet is consistently omitted; 
`a' refers to just one resolution is provided rather than the multiple-choice nonuniqueness facet being entertained. 
These are both self-consistent restrictions to make, leaving one with a rather simpler -- and solved -- problem of how to combine the nine local facets.  
The ensuing `A Local Resolution of the Problem of Time' 
has also been reformulated as \cite{APoT3, ABook, I, I, II, III, IV, V, VI, VII, VIII, IX, X, XI, XII, XIII, XIV} 
`A Local Theory of Background Independence'; see e.g. \cite{A64, A67, Giu09} for previous work on Background Independence.   

\m 

\n The classical part of this work \cite{I, II, III, IV, V, VI, VII, VIII, IX, X, XI, XII, XIII, XIV} 
requires just Mathematics along the lines of Lie's \cite{Lie, M08, Lee2} \cite{XIV}. 
The simpler parts of this use the following structures \cite{XIV}. 

\m

\n i) Lie derivatives \cite{Yano55, Yano70} to encode Relationalism.\footnote{See \cite{I, II, III, IV} 
for what Relationalism, Closure, observables, Spacetime Construction, and Refoliation Invariance entail; 
since the last two of these fall within the main subject of the current Article, outlines of them can also be found in Appendix \ref{GR-Examples}.}  

\m 

\n ii) Lie brackets assess Closure, whose end product (if successful) are {\it Lie algebraic structures}.
This means a portmanteau of Lie algebras \cite{Serre-Lie} and of Lie algebroids \cite{CM}: a distinction outlined in Sec \ref{LAS}, 
with the Dirac algebroid \cite{Dirac} (\ref{Mom,Mom}-\ref{Ham,Ham}) formed by GR's constraints as an example of the latter.  

\m 

\n iii) Further Lie brackets equations consistently define \cite{AObs} constrained notions of observables \cite{Dirac49, K93, ABook, VIII}. 
These Lie brackets equations can moreover be recast as \cite{ABook, VIII} first-order PDE systems.
These are amenable to Lie's flow method \cite{G63, John, Olver2, M08, Lee2, Olver} 
(for Finite Theories, or the functional DE analogue thereof for Field Theories including GR \cite{ABook, VIII}). 

\m

\n iv) Each theory's lattice of constraint algebraic substructures 
      induces a dual lattice of observables          subalgebras \cite{ABook, III}.

\m 

\n v)-viii) At the spacetime level, further Lie derivatives implementing Relationalism, Lie brackets assessing Closure, 
and associated Lie brackets based notions of observables occur \cite{APoT3, ABook, III, XI, XII}. 
Some scope for physically meaningful lattices of generator and observables subalgebras remains.  

\m  

\n See \cite{I, II, III, V, VI, VII, VIII, IX, XI, XII, XIII, XIV} for further details of basic occurrences of Lie's Mathematics in the classical part of 
A Local Resolution of the Problem of Time. 
The point of the current Article is, rather, to highlight that two Problem of Time facets involve some more advanced or modern `higher Lie Theory'. 

\m 

\n In Sec \ref{LR}, we outline {\it Lie Rigidity} for Lie algebras \cite{G64, NR64} and Lie algebroids \cite{CM}.  
In Sec \ref{LA}, we present `Lie's Algorithm' for Generator Closure: how Dirac's Algorithm for Constraint Closure generalizes to the general Lie bracket setting.  
In Sec \ref{LAR}, we reinterpret `passing families of theories \cite{RWR, AM13, ABook} 
through the Lie Algorithm', of which the Dirac Algorithm \cite{Dirac, HTBook, ABook, VII} 
subcase is better known \cite{RWR, AM13, IX}. 
This is a method used for Spacetime Construction (from space) and getting more structure from less structure assumed more generally. 
Our reinterpretation is as a {\it generator deformation approach} with the Dirac Rigidity subcase of Lie Rigidity then being seen to underlie the known results.   
In our concluding Sec \ref{Conclusion}, we point to such rigidity providing a partial cohomological (and thus global, algebraic-topological) selection principle 
for the Comparative Theory of Background Independence.

\m 

\n In Sec \ref{RIO}, we finally pose the universal (theory-independent) analogue of GR's Refoliation Invariance \cite{T73}, 
as {\it Reallocation of Intermediary-Object (RIO) Invariance} for general Lie Theory.   
This is an algebraic pentagon criterion: in evolving from an initial object to a final object, 
it amounts to switching which intermediary object one proceeds via amounting to at most a difference by an automorphism of the final object.  
Our concluding Sec \ref{Conclusion} also argues for this to also constitute a selection principle in the Comparative Theory of Background Independence.

\m 

\n In the current Article, we insist on putting the general Lie cases first -- Lie Algorithm, Lie Rigidity, Lie RIO-Invariance -- 
for all that one or both of the classical Dirac or GR versions of each of these are considerably better-known.
We compensate for this by providing Appendix \ref{GR-Examples} on the Dirac and GR versions. 
Appendix \ref{FOG-Example} then gives a Foundations of Geometry example of Lie Algorithm leading to a 2-pronged rigidity in the form of `top geometry': 
conformal versus projective; 
this demonstrates merit in extending from the Dirac Algorithm to the Lie Algorithm.

\section{Lie Algebraic Structures}\label{LAS}

\n Let us first introduce some notation. 
We use 
\be 
\mbox{\bf|[} \m \mbox{\bf ,} \m \mbox{\bf |]} 
\label{L-B}
\ee 
to denote Lie brackets, and 
\be 
\u{G}  \mbox ,  
\label{Gen}
\ee 
to denote infinitesimal algebraic structure generators; the underline here suppresses an in-general internal index.

\m  

\n{\bf Structure 1} If the generators close under the Lie brackets  
\be
\mbox{\bf |[} \u{G} \mbox{\bf ,} \,  \u{G}^{\prime} \mbox{\bf ]|}   \es   \u{\u{\u{\biG}}} \, \u{G}^{\prime\prime}      \m , 
\label{L-Algebra}
\ee 
for constants $\biG$, we have a {\it Lie algebra} 
\be
\FrL  \m , 
\ee 
these constants being termed {\it structure constants}.  

\m 

\n{\bf Structure 2} If the generators still close under Lie brackets
\be
\mbox{\bf |[} \u{G} \mbox{\bf ,} \, \u{G}^{\prime} \mbox{\bf ]| }   \es  \u{\u{\u{\biG}}}(\u{x}) \, \u{G}^{\prime\prime}    \m . 
\label{L-Algebroid}
\ee
but with {\it structure functions} $\biG(\u{x})$ instead of structure constants, we have a {\it Lie algebroid} (see e.g.\ \cite{G06} for an introductory exposition). 
We require this extension because of GR's Dirac algebroid (Appendix \ref{GR-Dirac}), 
kinematical quantization's \cite{M63, I84} modern reformulation \cite{Landsman} in terms of Lie algebroids, 
and a third reason given in Sec \ref{LAR}.

\section{Lie Rigidity}\label{LR}

{\bf Structure 1} This starts at the level of generator deformations, which we denote by  
\be 
\u{G} \longrightarrow \u{G}_{\alpha}  \es  \u{G} + \alpha \, \phi   \m . 
\label{def}
\ee
$\alpha$ is here in general a multi-index, so one has in general the corresponding multi-index inner product with an equally multi-indexed set of functions $\phi$. 
These deformed generators can a priori be viewed as 1) terminating at linear order in $\alpha$, 
                                                 or 2) as infinitesimal            in $\alpha$ so that $\alpha^2 = 0$ by fiat. 
In fact our interpretation in the next section's setting is the former, since each of our $\alpha$ is typically a priori real-valued.

\m 

\n{\bf Remark 1} Gerstenhaber \cite{G64} classifies types of notion of deformation. 
(\ref{def}) is a real-geometric notion of deformation \cite{G64}. 
It is moreover also either an infinitesimal deformation for 1) or a first-order truncated deformation for 2).   
Gerstenhaber initially works in the setting of associative algebras; Nijenhuis and Richardson \cite{NR64} then specialize to the case of Lie algebras.  

\m 

\n{\bf Structure 2} Gerstenhaber\cite{G64} additionally attributes local stability under deformations to rigid algebraic structures. 
He also places a cohomological underpinning on rigidity results, 
which Nijenhuis and Richardson \cite{NR66} again specialize to the Lie algebra case as follows.
\be 
\mH^2(\FrL, \FrL) = 0   \m .  
\ee 
diagnoses rigidity.  
This $\mH^2$ cohomology group consists of the quotient of the group 
\be 
\mZ_2(\FrL, \FrL)
\ee 
of Lie algebra 2-cocycles:  
\be 
\phi : \FrL \times \FrL \longrightarrow \FrL 
\ee
such that 
\be 
\phi(\mbox{\bf |[} X \mbox{\bf ,} \, Y \mbox{\bf ]|} , \, Z) + \mbox{cycles} - \mbox{anticycles}  \es  0
\ee 
by the group 
\be 
\mB_2(\FrL)
\ee 
of coboundaries: linear maps  
\be 
\psi: \FrL \longrightarrow \FrL 
\ee 
such that 
\be 
\phi(X, Y)  \es  (d_1 \psi)(X, Y)  \es   \psi(\mbox{\bf |[}X \mbox{\bf ,} \, Y \mbox{\bf ]|}) 
                                       - \mbox{\bf |[} X \mbox{\bf ,} \, \psi(Y)\mbox{\bf ]|} 
									   + \mbox{\bf |[} Y \mbox{\bf ,} \, \psi(X) \mbox{\bf ]|}  
\ee 
for 1-coboundary $d_1$.  

\m 

\n{\bf Remark 2} The corresponding $\mH^1$ group is itself tied to the simpler matter of Lie group automorphisms.

\m 

\n{\bf Remark 3} By evoking cohomology, this takes us beyond Lie's Mathematics into the terrain of e.g.\ Poincar\'{e}'s, \v{C}ech's, and de Rham's Mathematics 
(and further categorical abstraction as per the 1950s and 1960s; see \cite{BT82, MT, Steenrod, Hatcher} for reviews) 
and further Lie-theoretic specifics from the 1960s \cite{G64, NR64, G66, NR66, G68}.

\m 

\n{\bf Remark 4} `Deformation' is meant here in the same kind of sense as in `deformation quantization' \cite{L78, S98, Kontsevich}.  
In this way, such deformations are a matter of some familiarity to Theoretical and Mathematical Physics.  
The current Article's application is moreover a clearly distinct -- entirely classical -- application of deformation.

\m

\n{\bf Remark 5} The current Article's application extending to algebroids moreover takes one out of Gerstenhaber's original setting \cite{G64} 
for deformations and rigidities -- algebras -- for which Nijenhuis and Richardson \cite{NR64} providing further Lie algebra specifics. 
However, e.g.\ Crainic and Moerdijk \cite{CM} subsequently considered matters of rigidity for algebroids, 
so this case remains posed and with some significant results.

\section{The Lie Algorithm}\label{LA}

\n{\bf Structure 1} Suppose we are given an incipient set of algebraic generators 
\be 
\u{G}(\biB)  \mbox ,  
\ee 
where the $\biB$ are `base objects', e.g.\ configurations $\biQ$, that our generators are built out of.  
\n We form their Lie brackets 
\be 
\mbox{\bf|[} \u{G} \mbox{\bf ,} \, \u{G}^{\prime} \mbox{\bf |]} 
\ee 
to assess whether they close, either by themselves or through further algebraic objects arising in the process. 

\m 

\n Fundamental Physics requires this working to reside within the realm of {\it Lie brackets algebraic structures}. 
This is by Appendix \ref{GR-Dirac}'s Lie algebroid example as well as due to quantum-level reinterpretation \cite{Landsman} of 
Mackey-type kinematical quantization \cite{M63, I84} in terms of algebroids.
We thus require the base-object version of (\ref{L-Algebroid}), 
\be
\mbox{\bf |[} \u{G} \mbox{\bf ,} \, \u{G}^{\prime} \mbox{\bf ]| }   \es  \u{\u{\u{\biG}}}(B) \, \u{G}^{\prime\prime}    \m . 
\ee
\n The particular situation we consider is placing an incipient set of generators into a `Lie Algorithm'. 

\m 

\n Six types of equation can arise in the process, 
the first five of which correspond to the five familiar from the classical Dirac Algorithm \cite{Dirac, HTBook}. 
See \c{XIV} for a case-by case analysis of motivation and scope.  
                                              
\m 

\n{\bf Case 1)} An identity equation (like 0 = 0). 

\m 

\n{\bf Case 2)} An inconsistency (like 0 = 1).

\m 

\n{\bf Case 3)} An extra generator is found.  

\m 

\n{\bf Remark 1} Such extra generators could cause an enlarged algebraic structure, 
or a cascade of further terms reducing a candidate theory to triviality or inconsistency \cite{ABook, IX}.

\m 

\n{\bf Case 4)} A {\it need to rebracket} is uncovered, 
by {\it Lie irreducibly-second-class objects} 
requiring passage to the {\it Lie-Dirac bracket} 
(arbitrary Lie-theoretic generalization \cite{VII, XIV} of Dirac-second-class constraints and Dirac brackets \cite{Dirac}). 

\m 

\n{\bf Case 5)} Suppose there is involvement of a procedure involving appending by auxiliaries 
(of which the Dirac Algorithm \cite{Dirac, ABook} is probably the best-known example). 
Then a priori free auxiliaries can also become specified by a further kind of relation arising from the algebraic structure. 
We term these `{\it Lie specifier equations}' \cite{VII, XIV} (with, matchingly `Dirac specifier equations' constituting the most well-known subcase).

\m 

\n{\bf Case 6)} A topological obstruction; this could be a `hard' enough feature or one characteristic of topology.   
This case is more commonly encountered at the quantum level, consisting of phenomena along the lines of anomalies \cite{Bertlmann}.

\section{How widespread is Lie Rigidity within the Lie Algorithm?}\label{LAR}

\n{\bf Structure 1} We next follow \cite{RWR, AM13, ABook, IX, XIV} in putting entire families of generators into the Lie Algorithm.   

\m 

\n We now furthermore identify this procedure as deformation of (some of the) input generators. 
At the level of the brackets algebraic structure itself, this sends 
\be 
\mbox{\bf |[} \u{G}\mbox{\bf ,} \, \u{G}^{\prime} \mbox{\bf ]|}  \es  \u{\u{\u{\biG}}} \, \u{G}^{\prime\prime}  
\ee 
to
\be   
\mbox{\bf |[} \u{G}\mbox{}_{\alpha}, \, \u{G}^{\prime}\mbox{}_{\alpha} \mbox{\bf ]|}   \es  \u{\u{\u{\biG}}}(\alpha) \, \u{G}^{\prime\prime}\mbox{}_{\alpha}  +  
                                                                                            \u{\u{\u{\biH}}}(\alpha) \, \u{H}\mbox{}_{\alpha}                 +  
																	                        f(\alpha) \u{\u{\Theta}} \, \mbox{}_{\alpha}                        \m .
\label{alpha-Alg}
\ee
\n{\bf Remark 1} These $\u{H}\mbox{}_{\alpha}$ are integrabilities of the $\u{G}\mbox{}_{\alpha}$.  

\m 

\n{\bf Remark 2} In our mathematical arena of interest, (\ref{alpha-Alg}) includes in principle the eventuality that 
a Lie algebra's structure constants $\biG$ deform to a Lie algebroid's structure functions  $\biG(\alpha)$, 
or extend thereto        (i.e.\ the $\biH(\alpha)$ could be structure functions even if the $\biG(\alpha)$ are not).  
The latter algebroid feature takes one out of Gerstenhaber's original algebraic setting \cite{G64}. 
I.e.\ deformation itself gives a third reason (to Sec \ref{LAS}'s two) for involvement of Lie algebroids.  

\m 

\n{\bf Remark 3} By the final zeroth-order right-hand-side term, topological obstructions, e.g.\ along the lines of anomalies can enter proceedings.  

\m 

\n{\bf Remark 4} The $\biG(\alpha)$,  $\biH(\alpha)$ and $f(\alpha)$ can moreover have strongly vanishing roots, 
i.e.\ particular values of $\alpha$ for which these terms disappear. 
On some occasions, this is capable of picking out special cases that remain free of topological obstructions, 
                                                                                    integrabilities, 
																					need for rebracketing, 
																					specifers, or 
																					structure functions. 

\m 

\n{\bf Remark 5} If integrabilities occur, one needs to consider the further self brackets 
\be 
\mbox{\bf |[} \u{H}\mbox{}_{\alpha} \mbox{\bf ,} \,  \u{H}\mbox{}_{\alpha} \mbox{\bf ]|}  
\ee 
and mutual brackets 
\be 
\mbox{\bf |[} \u{G}\mbox{}_{\alpha} \mbox{\bf ,} \,  \u{H}\mbox{}_{\alpha} \mbox{\bf ]|}
\ee 
to see if topological obstructions, 
          further integrabilities, 
		  distinct realizations of structure functions, 
         (or further strong vanishings preventing whichever of these) arise.

\m 

\n One continues forming brackets until if and when one reaches an iteration at which no new objects arise. 

\m 

\n In all examples considered in the current article, there is either a non-strongly vanishing obstruction or a theory-killing cascade cofactor.
This is avoided by picking a strongly-vanishing root, rigidly returning a theory of interest (or a contraction thereof in some cases). 

\m

\n{\bf Remark 6} How hard and final integrabilities and obstructions are can vary.  
For instance, integrabilities can cascade to eat up all degrees of freedom, leaving a candidate theory with none, or beyond, rendering the candidate inconsistent.   

\m 

\n{\bf Remark 7}  The outcome of putting deformed Lie generators through the Lie Algorithm leaves us facing the question of when `anything goes' and when 
just one (or very few) sharp possibilities occur. 
This bears some relation to which Lie theories are rigid under deformations. 

\m 

\n Some examples of this (Appendix \ref{Sp-C}) moreover also appear to realize that contracted \cite{Gilmore} limits of a given algebraic structure remain consistent, 
but have to be encoded separately from the uncontracted version.
\cite{L67} is the first known instance of contractions being treated alongside rigidities.  

\m 

\n{\bf Remark 8} We need to work with maximally general deformations to ensure that rigidity results do not disappear upon considering furtherly general deformations.

\m 

\n{\bf Remark 9} We would do well to use further examples to assess how typical is it 
for contractions of an algebraic structure to accompany uncontracted versions thereof as other means of attaining consistency in the Lie Algorithm.  

\m 

\n{\bf Remark 10} Evoking rigidity takes one away from local considerations, 
at the level of the Lie algebraic structure of constraints, or of spacetime automorphism generators.
It gives a cohomological underpinning to each corresponding aspect of Comparative Background Independence \cite{A-Killing, A-Cpct, A-CBI}.  
I.e.\ for which configuration space 
\be 
\FrQ
\ee 
and which 
\be 
Aut(\FrQ, \sigma)
\ee 
do these $\mH^2$ cohomology groups vanish, leaving one with a rigid recovery of the spacetime version of the structure? 
This constitutes a selection principle among Backgound Independent theories.

\m 

\n[Comparative Background Independence in good part \cite{ASoS, A-Killing, A-Cpct, A-CBI} concerns the relational configuration spaces 
\be 
\frac{\FrQ}{Aut(\FrQ, \sigma)}     \m , 
\ee 
where $\sigma$ is the level of mathematical structure included, e.g. metric geometry or affine geometry.]

\m 

\n Or similarly, as a selection principle at the level of recovering one spatial level of structure from a lesser level of spatial structure 
(see \cite{IX} for examples). 
Or yet again, now at the level of recovering one spacetime level of structure from a lesser level of spacetime structure.

\section{Which theories have Reallocation of Intermediary-Object Invariance?}\label{RIO}
%
{            \begin{figure}[!ht]
\centering
\includegraphics[width=0.3\textwidth]{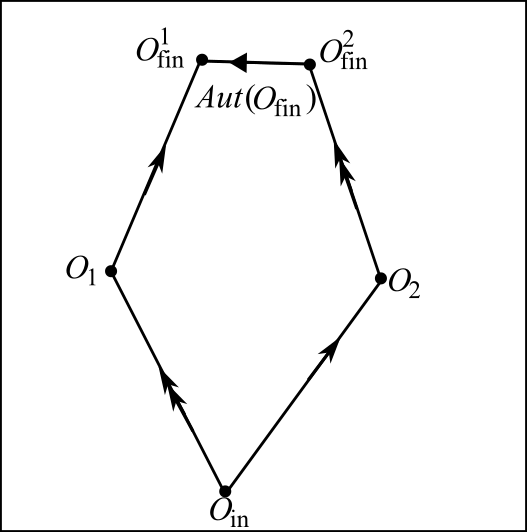}
\caption[Text der im Bilderverzeichnis auftaucht]{ \footnotesize{Commuting pentagon from $O_{\si\sn}$ to $O_{\sf\si\sn}$ via 
two distinct allocations of intermediary objects, $O_1$ and $O_2$.}}
\label{Pentagon}\end{figure}            }

\n{\bf Example 1} A well-known case is Refoliation Invariance \cite{T73, K92, I93, ABook, XII} in GR [posed in Fig 2.b) and resolved in Fig 2.c)]. 

\m

\n This generalizes to the following universally poseable, if not necessarily realizable, structure.   

\m  

\n{\bf Definition 1} {\it Reallocation of Intermediary-Object (RIO)} is the commuting-pentagon property depicted in Figure 1.  
In more detail, it is a {\it commuting square}, corresponding to moving from an initial object $O_{\si\sn}$ to a final object $O_{\sf\si\sn}$ 
via two different intermediaries $O_{1}$ and $O_{2}$, {\it up to some automorphism of the final object}, 
\be 
Aut(O_{\sf\si\sn})  \m ,  
\ee 
relating the outcomes of proceeding via $O_{1}$ and via $O_{2}$.  
This automorphism constitutes the fifth side of the pentagon.  

\m 

\n{\bf Remark 1} One or both of $O_1$ and $O_2$ can be replaced with distinct arbitrary intermediate objects. 

\m 

\n{\bf Remark 2} Some theories will obey this property, and some will not (see Appendix \ref{Refol-Inv} and the Conclusion for examples). 
RIO invariance thus also has the status of a selection principle.

\section{Conclusion}\label{Conclusion}

The current Article points to some algebraic content of the Comparative Theory of Background Independence. 
Namely 1) a class of Lie algebaic structures that rigidly resist deformation and 2) a type of commutative pentagon diagram. 
1) covers both Spacetime Constructability from space 
          and  recovery from fewer levels of structure known. 
2) refers to Reallocation of Intermediary-Object (RIO) Invariance.   
These properties have already quite long been remarked upon in GR, for which 
1) includes Dirac Rigidity of GR recovering spacetime from just space \cite{RWR}, whereas 
2) is manifested as Refoliation Invariance \cite{T73}. 
Each of these resolves one of GR's Problem of Time facets, 
corresponding to two ways \cite{K92, I93, ABook, III, IX, XII} in which GR succeeds in being Background Independent.
Simpler parts of Lie's Mathematics perform the same function for five further Problem of Time facets \cite{ABook, I, II, III, V, VI, VII, VIII, X}, 
thus encoding five further Background Independence aspects. 

\m 

\n By tying Spacetime Constructability's families of constraints (or underlying families of relational actions \cite{IX} encoding those constraints) 
to deformations of generators, we moreover introduce the Gerstenhaber--Nijenhuis--Richardson cohomological characterization of Lie Algebra Rigidity 
into the Problem of Time and Background Independence literature. 
One in fact needs a Lie Algebroid Rigidity extension of this cohomological characterization due to GR's Dirac Algebroid and generalizations 
\cite{PVM, Bojo12, Bojo16} as well as due to kinematical quantization's \cite{M63, I84} modern reformulation \cite{Landsman} in terms of Lie algebroids.  
Crainic and Moerdijk \cite{CM} have provided some results about extensions of Lie Rigidity to algebroids.    

\m 

\n Such cohomological characterization constitutes a further aspect of the Global Problems of Time\footnote{See 
Epilogues II.B and III.B of \cite{ABook} for further distinct Global Problems of Time, 
and Epilogue II.C and \cite{A-Killing, A-Cpct, A-CBI} for further global aspects of Comparative Background Independence specifically.}  
that is moreover a useful criterion in the theory of Comparative Background Independence.  
I.e.\ one would expect only some theories with candidate status of Background Independent theories to be rigid in this manner.  
The Comparative Theory of Background Independence is thereby not only an algebraic venture, but more specifically a venture in Algebraic Topology.    

\m 

\n For now, this is being considered \cite{IX} in the context of differential-geometric theories with variable levels of structure, 
which we term `moderate' Comparative Background Independence.
This is as opposed to Background Independence to any level of structure in the Standard Foundational System of Mathematics 
(\cite{I89-I91, ASoS}, Epilogue II.C of \cite{ABook}) or Background Independence of {\sl any} structure (\cite{I-Cat}, and Epilogue III.C of \cite{ABook}).  
These more general cases require further work because of loss of contact with Lie's familiar work 
or at least the need to generalize this work to apply in increasingly wide contexts.  

\m 

\n Canonical GR's possession of Refoliation Invariance is key to fleets of observers 
-- each accelerating differently and thus passing through a distinct sequence of local frames -- 
being able to reach observational concordance (Appendix \ref{Refol-Inv}).  
Refoliation Invariance, or a suitable generalization, is thus a desirable feature for a Background Independent theory to possess.  
RIO Invariance is such a generalization that is poseable for any candidate theory. 
It is moreover expected that only some candidate theories would obey this property, by which RIO Invariance is a selection principle. 
For instance, fixed-frame or privileged-frame (sometimes alias fixed-slicing or privileged-slicing) theories do not satisfy RIO Invariance.  

\m 

\n Let us finally note that everything outlined in the current Article can furthermore be 
{\sl posed} for Supersymmetric Theories including for Supergravity in place of GR \cite{PVM}.
In the case of rigidities, this field was indeed essentially ab initio formulated for graded (Lie) algebras \cite{G66, NR66, G68}: 
the structures subsequently used in Supersymmetric Theories.  
As such, a supersymmetric (or graded) generalization of the current Article is readily available.  
 
\m 

\n{\bf Acknowledgments}

\m 

\n I thank Paolo Vargas--Moniz, whose posing of the Supergravity version of Refoliation Invariance \cite{PVM} 
is a significant, if as yet unanswered, precursor of what the current brief article argues to be the heart of Comparative Background Independence, 
and various friends for support.    

\begin{appendices}

\section{GR examples}\label{GR-Examples}

\subsection{GR constraints and the Dirac algebroid they form}\label{GR-Dirac}

\n{\bf Structure 1} GR's constraints \cite{ADM} are, firstly,   
\beq
\mbox{\it GR Hamiltonian constraint } \m  \scH  \:=  {||\mp||_{\sbN}}^2  - \overline{{\cal R} - 2 \, \slLambda}   
                                                \es  \mN_{ijkl}\mp^{ij}\mp^{kl} - \overline{{\cal R} - 2 \, \slLambda} 
                                                \es  0                                                                        \m , 
\label{Hamm}
\eeq 
where $\bh$ with components $\mh_{ij}$ is the spatial 3-metric 
with determinant $\mh$, Ricci scalar ${\cal R}$ 
and conjugate momentum $\bp$ with components $\mp^{ij}$.  
The overline denotes densitization, i.e.\ multiplication by $\sqrt{\mh}$.  
\be
\bN \m \mbox{ with components } \m \mN_{abcd}  \es  \frac{1}{\sqrt{\mh}} \{  \mh_{ac}\mh_{bd} - \half \mh_{ab}\mh_{cd}  \}                           \m , 
\ee
is the DeWitt supermetric \cite{DeWitt67}, and $\slLambda$ is the cosmological constant.   

\m 

\n Secondly, the 
\beq
\mbox{\it GR momentum constraint } \m \u{\sbcM} :=   - 2 \, \u{\bcalD} \, \u{\u{\bp}} = 0 \m \mbox{ with components } \m 
                                      \scM_{i}  \es  - 2 \, {\cal D}_{j} {\mp^{j}}_{i} 
                                                \es                   0                                                       \m ,  
\label{Momm}
\eeq
where $\bcalD$ with components ${\cal D}_i$ is the spatial covariant derivative.  

\m 

\n The coordinate-independent presentations above are the ones used in the outline account below.  

\m 

\n{\bf Structure 2} The Dirac algebroid \cite{Dirac51, Dirac58, Dirac} formed by GR's constraints takes the coordinate-independent form 
\be
\mbox{\bf \{} ( \u{\bscM}  \,  |  \,  \u{\bupxi}  ) \mbox{\bf ,} \, (  \u{\bscM}  \,  |  \,  \u{\bupchi}  )  \mbox{\bf \}}   \es   (  \u{\bscM}    \,  |  \, \, \u{[ \bupxi, \bupchi ]}  )    \m ,
\label{Mom,Mom}
\ee
\be
\mbox{\bf \{} (  \scH    \,  |  \,  \upmu  ) \mbox{\bf ,} \, (  \u{\bscM}  \, | \,  \u{\bupxi}  ) \mbox{\bf \}}   \es   (  \pounds_{\u{\sbupxi}} \scH  \,  |  \,  \upmu  )  \m , 
\label{Ham,Mom}
\ee
\be 
\mbox{\bf \{} (  \scH    \,  |  \,  \upzeta  ) \mbox{\bf ,} \, (  \scH  \, | \,  \upomega  ) \mbox{\bf \}}   \es   
( \u{\bscM} \cdot \u{\u{\bh}}^{-1} \cdot \, | \, \upzeta \, \overleftrightarrow{\u{\bpa}} \upomega ) \m . 
\label{Ham,Ham}
\ee
$( \m | \m )$ is here the integral-over-space functional inner product, 
$[ \m , \m ]$, the differential-geometric commutator Lie bracket, 
and $\bupxi$, $\bupchi$, $\zeta$ and $\upomega$ are smearing functions. 
Such `multiplication by a test function' serves to render rigourous a wider range of 
     `distributional' manipulations provided that these occur under an integral sign.  

\m 

\n If one starts with GR's constraints and places them into the Dirac Algorithm \cite{Dirac}, one finds this Dirac algebroid. 
This amounts to consistency, with $\scH$ and $\u{\scM}$ having the status of first-class constraints \cite{Dirac}. 
From the structure functions in the last equation's right-hand-side, the Dirac algebroid is indeed an algebroid \cite{BojoBook}.

\subsection{Refoliation Invariance}\label{Refol-Inv}

\n{\bf Definition 1} {\it Refoliation Invariance} \cite{T73} is encapsulated by evolving via each of Fig \ref{Refol-Inv-Fig}.b)'s red and purple hypersurfaces 
giving the same physical answer as regards the final hypersurface. 

\m 

\n{\bf Remark 1} GR spacetime is thus not just a single strutting together of spaces like Newtonian space-time is.  
GR spacetime manages, rather, to be many such struttings at once in a physically mutually consistent manner, as per Fig \ref{Refol-Inv-Fig}.c).
Indeed, this is how GR is able to encode consistently the experiences of fleets of observers moving in whichever way they please. 

\m 

\n{\bf Remark 2} It is the plurarity of intermediary hypersurfaces that requires algebroid structure to encode.  

\m 

\n{\bf Remark 3} The Refoliation Invariance result comes about by the third relation in GR's Dirac algebroid (\ref{Ham,Ham}) 
being none other than a local algebraic formulation of Refoliation Invariance (Fig \ref{Refol-Inv-Fig}.c). $\Box$  

\m 

\n{\bf Remark 4} This is accompanied by two other commuting pentagons: Fig 2.d-e), corresponding to the first two equations in the Dirac algebroid.  
%
{            \begin{figure}[!ht]
\centering
\includegraphics[width=1.0\textwidth]{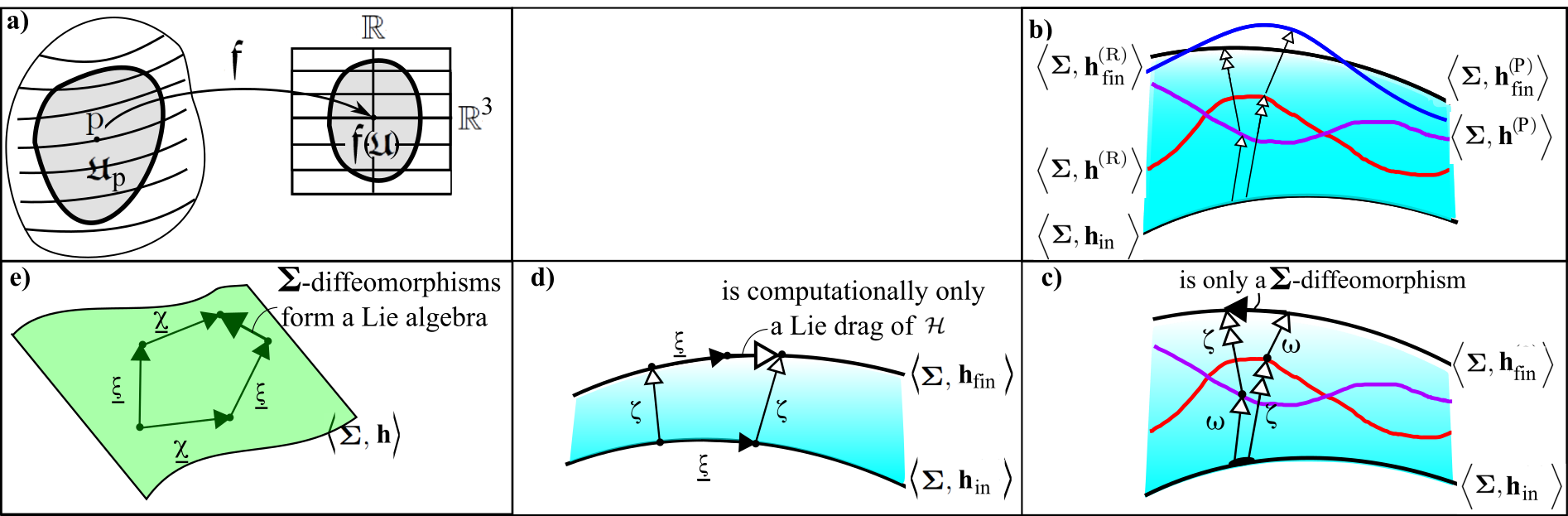}
\caption[Text der im Bilderverzeichnis auftaucht]{ \footnotesize{a) Illustrating the nature of {\it foliation} $\sFrf$: 
a decorated (or, more precisely, rigged) version of the standard Differential-Geometric definition of chart on a manifold $\FrM$.  

\m 

\n b) Posing Refoliation Invariance: is going from spatial hypersurfaces in to fin 
via the red (R) intermediary hypersurface being physically the same as going via the purple (P) intermediary surface? 
If so, the blue and black hypersurfaces would coincide. 

\m 

\n c) For GR, however, the Dirac algebroid's eq. (\ref{Ham,Ham}) gives that the black and blue hypersurfaces can at most differ by a spatial diffeomorphism, 
and so must coincide as the same geometrical entity (R)-fin = fin = (P)-fin
This can be seen as the diffeomorphism establishing that an a priori distinct square of maps is in fact commutative.

\m 

\n GR's Dirac algebroid's supporting relations' (\ref{Ham,Mom}) and (\ref{Mom,Mom}) geometrical meanings are depicted in d) and e). } }
\label{Refol-Inv-Fig}\end{figure}            }

\subsection{Spacetime Constructability}\label{Sp-C}

\n{\bf Structure 1} Suppose one assumes less structure than is present in GR's notion of spacetime, 
one needs to recover it from what structure is assumed (at least in a suitable limit).  
We use `Spacetime Constructability'     for the Background Independence aspect,  
                              `Spacetime Construction Problem' for the Problem of Time facet in cases in which this is blocked, and 
                              `Spacetime Construction'         for corresponding strategies.  
Already at the classical level, Spacetime Constructability can moreover be considered in two logically independent senses.

\m 

\n {\bf Sense A)} Spacetime Constructability from space. 

\m 

\n {\bf Sense B)} Spacetime Constructability from assuming fewer levels of mathematical structure (see \cite{IX} for examples).

\m 

\n{\bf Structure 2} We consider the family of trial constraints   
\be 
\scH_{\st\sr\si\sa\sll}  \es  \scH_{x,y,a,b}  
                         \:=  {||\bp||_{\sbN^{x,y}}}^2 - \overline{a \, {\cal R} + b}  
						 \es  0                                                                 \m ,
						 \label{H-trial}
\ee 
for constants $x$, $y$, $a$, $b$, and where 
\be 
\bN^{x,y} \m \mbox{ has components } \m  
\mN_{abcd}^{x,y} := \frac{y}{\sqrt{h}}\left\{  \mh_{ac}\mh_{bd} -  \frac{x}{2} \,\mh_{ab}\mh_{cd} \right\}
\ee 
and with GR's DeWitt supermetric $\bN$ \cite{DeWitt67} corresponding to the $x = 1$ case.  
We consider the family of constraints (\ref{H-trial}) alongside the standard momentum constraint $\u{\scM}$.  

\m 

\n{\bf Structure 3} The self-bracket of $\scH_{\st\sr\si\sa\sll}$ now picks up an extra additive term as an 
{\it obstruction term to having a brackets algebraic structure} \cite{RWR, San, Phan} 
\be
2 \times a \times y \times \{ 1 - x \} \times (    \bcalD \, \mp \, | \, \xi \, \overleftrightarrow{\bpa} \zeta    )   \m .  
\label{4-Factors}
\ee 
This has four factors. 
Each of these being zero providing a different way in which to attain consistency. 
GR follows \cite{RWR} from setting $x = 1$ corresponding to the third factor being satisfied.
  
\m 

\n{\bf Remark 1} The first factor in (\ref{4-Factors}) vanishing corresponds to Strong Gravity; see \cite{I76, Henneaux79} for Strong Gravity in general, and
                                                                                                    \cite{OM02, San} in the consistent brackets algebra context. 

\m 

\n The second factor corresponds to Galileo--Riemann Geometrostatics, first featuring in the consistent brackets algebraic structure context in \cite{Lan}.  

\m 

\n The fourth, now merely weakly vanishing factor, corresponds to a privileged constant mean curvature (CMC) foliation.
For this, the GR constraints decouple: methodology already well-established in the GR Initial-Value Problem \cite{York73}.  

\m

\n{\bf Remark 2} The conceptual type of (\ref{4-Factors}) is thus  
\be 
(\mbox{Galileo--Riemann geometrostatics})           \times 
(\mbox{GR's Geometrodynamics' DeWitt value})        \times 
(\mbox{Strong Gravity})                             \times 
[\mbox{CMC foliation}]                                        \m ,    
\ee
where round brackets denote strongly-vanishing factors 
 and square brackets for      weakly-vanishing factors.  

\m 

\n{\bf Structure 4} Upon including minimally-coupled matter terms as well, a second obstruction term occurs \cite{RWR, AM13, ABook}.
Its factors include the ambiguity Einstein faced of whether the universal Relativity is locally Galilean \cite{Lan} or Lorentzian. 
This ambiguity is based on whether the fundamental universal propagation speed $c$ (often referred to as `speed of light') takes an infinite value, 
or a fixed-finite one $c_{\su\sn\si}$. 
Carrollian Relativity \cite{San}, corresponding to zero such speed, now features as a third option as well, corresponding to ultralocal matter \cite{Klauder70}.   

\m 

\n This ambiguity is now moreover realized in the {\sl explicit mathematical form} of a string of numerical cofactors 
of what would otherwise be an {\sl obstruction term to having a brackets algebraic structure of constraints}.  
In particular, the GR spacetime solution to the first obstruction term 
is now accompanied by second obstruction term's condition that {\sl the locally-Lorentzian relativity of SR is obligatory}.  
This can be viewed as all minimally-coupled matter sharing the same null cone \cite{RWR} 
because each matter field is separately obliged to share {\sl Gravity}'s null cone.  

\m 

\n{\bf Remark 3} The conceptual type of this further obstruction term is thus of the form 
\be 
(\mbox{$c = \infty$ contraction})           \times 
(\mbox{$c = c_{\su\sn\si}$, fixed finite})  \times 
(\mbox{$c = 0$ contraction})                \times 
[\mbox{matter variation term}]                                 \m , 
\ee
i.e.\ 
\be 
(\mbox{Galilean Relativity})                \times 
(\mbox{Locally Lorentzian Relativity: SR})  \times 
(\mbox{Carollian Relativity})               \times 
[\mbox{matter variation term}]                                  \m .  
\ee
\n{\bf Remark 4} The combination of GR's particular $\scH$ alongside local Lorentzian Relativity and 
embeddability into GR spacetime thus arises as one of very few consistent possibilities, the others of which are at least of conceptual interest.  

\m 

\n{\bf Remark 5} Strong Gravity moreover pairs as a natural setting for Carrollian Relativity, 
and Galileo--Riemann Geometrostatics with a generalization of Galilean Relativity.  

\m 

\n{\bf Remark 6} The above strings of numerical factors moreover arise {\sl from the Dirac Algorithm} as the choice of factors among which a strong one needs 
to vanish in order to avoid the constraint algebroid picking up an obstruction term.
(35-36) moreover arise in a substantially different from how Einstein's dichotomy between universal local Galilean or Lorentzian Relativity did;   
{\sl now mere algebra gives us this dichotomy} (within a larger ambiguity).  

\m 

\n{\bf Remark 7} For ease of presentation, the current Appendix's GR results are given as piecemeal resolutions to single Problem of Time facets. 
See \cite{VII, XII, IX} for more detailed exposition of how to combine these with other Problem of Time facets' resolutions.
Use Chapter 34 of \cite{ABook} while \cite{XII} remains unavailable.

\section{Foundations of Geometry example of Lie Algorithm}\label{FOG-Example}

{\bf Structure 1} The general (bosonic vectorial) quadratic generator in $\geq 2$-$d$ is given by the following 2-parameter family ansatz: 
\be 
\u{Q}^{\st\sr\si\sa\sll}_{\mu, \nu}  \:=  \mu \, ||x||^2\u{\pa} + \nu \, \u{x} \, (\u{x} \cdot \u{\nabla} )                             \m .  
\ee 
This arises from consideration of the general fourth-order isotropic tensor contracted into a symmetric object $x^a x^b$ (flat spatial indices).    

\m 

\n{\bf Theorem} \cite{A-Brackets} For $d \geq 2$, $\u{Q}^{\st\sr\si\sa\sll}_{\mu, \nu}$ self-closes only if $\mu \, (2 \, \mu + \nu) = 0$.

\m 

\n{\bf Remark 1} This arises as a strong vanishing to avoid an obstructory cofactor, schematically 
\be 
(\mbox{top geometry = projective}) \times 
(\mbox{top geometry = conformal})  \times  
[\mbox{obstuction term}]                      \m .  
\ee
\n{\bf Remark 2} The first factor vanishing is a recovery of the special-projective generator 
\be 
\u{Q} =  \u{x} \, (\u{x} \cdot \u{\nabla} )            \m , 
\ee 
whereas the second is a recovery of the special-conformal generator            
\be 
\u{K} =  ||x||^2 \u{\pa} - 2 \, \u{x} \, (\u{x} \cdot \u{\nabla})  \m . 
\ee
\n{\bf Remark 3} In this manner, the conformal versus projective alternative for flat-space top geometry is recovered.  			   
{\it Q.E.D.} that Lie brackets rigidity, outside of Dirac's Poisson brackets rigidity, is a realized phenomenon, 
and extending the Dirac Algorithm to a `Lie Algorithm' has merit.  

\end{appendices}



\begin{thebibliography}{99}

\footnotesize

%
\bibitem{Lie}                 S. Lie and F. Engel, {\it Theory of Transformation Groups} Vols 1 to 3 (Teubner, Leipzig 1888-1893); 
                              for an English translation with modern commentary of Volume I, see J. Merker (Springer, Berlin 2015), arXiv:1003.3202.   
  
\bibitem{Dirac49}             P.A.M. Dirac, ``Forms of Relativistic Dynamics", Rev. Mod. Phys. {\bf 21} 392 (1949).

\bibitem{Dirac51}             P.A.M. Dirac, ``The Hamiltonian Form of Field Dynamics", Canad. J. Math. {\bf 3} 1 (1951).

\bibitem{Steenrod}            S. Eilenberg and N. Steenrod.	{\it Foundations of Algebraic Topology} (Princeton University Press 1952).

\bibitem{Yano55}              K. Yano, ``Theory of Lie Derivatives and its Applications (North-Holland, Amsterdam 1955).   
				
\bibitem{Dirac58}             P.A.M. Dirac, ``The Theory of Gravitation in Hamiltonian Form", 
                              {\it Proceedings of the Royal Society of London} {\bf A 246} 333 (1958).

%
\bibitem{ADM}                 R. Arnowitt, S. Deser and C.W. Misner, ``The Dynamics of General Relativity", 
                              in {\it Gravitation: An Introduction to Current Research} ed. L. Witten (Wiley, New York 1962), arXiv:gr-qc/0405109. 
	
\bibitem{G63}                 H.W. Guggenheimer, {\it Differential Geometry} (McGraw--Hill, New York 1963, reprinted by Dover, New York 1977).  		

\bibitem{M63}                 G. Mackey, {\it Mathematical Foundations of Quantum Mechanics} (Benjamin, New York 1963).
 	 		
\bibitem{A64}                 J.L. Anderson, ``Relativity Principles and the Role of Coordinates in Physics.", in {\it Gravitation and Relativity} 
                              ed. H-Y. Chiu and W.F. Hoffmann p.\  175 (Benjamin, New York 1964).   
							  			
\bibitem{Dirac}               P.A.M. Dirac, {\it Lectures on Quantum Mechanics} (Yeshiva University, New York 1964). 

\bibitem{G64}                 M. Gerstenhaber, ``On the Deformation of Rings and Algebras", Ann. Math. {\bf 79} 59 (1964).  

\bibitem{NR64}                A. Nijenhuis and R. Richardson,``Cohomology and Deformations of Algebraic Structures" Bull. Amer. Math. {\bf 70} 406 (1964).  
                              
\bibitem{Serre-Lie}           J.-P. Serre, {\it Lie Algebras and Lie Groups} (Benjamin, New York 1965); 

                                           {\it Complex Semisimple Lie Algebras} (Springer, New York 1966).
 
\bibitem{G66}                 M. Gerstenhaber, ``On the Deformation of Rings and Algebras. II.", Ann. Math. {\bf 84} 1 (1966).  

\bibitem{NR66}                A. Nijenhuis and R. Richardson, ``Cohomology and Deformations in Graded Lie Algebras" Bull. Amer. Math. {\bf 72} 406 (1966).  

\bibitem{A67}                 J.L. Anderson, {\it Principles of Relativity Physics} (Academic Press, New York 1967).  

\bibitem{L67}                 M. Levy-Nahas ``Deformation and Contraction of Lie Algebras", J. Math. Phys. {\bf 8} 1211 (1967). 

\bibitem{Battelle}            J.A. Wheeler, in {\it Battelle Rencontres: 1967 Lectures in Mathematics and Physics} 
                              ed. C. DeWitt and J.A. Wheeler (Benjamin, New York 1968).  

\bibitem{DeWitt67}            B.S. DeWitt, ``Quantum Theory of Gravity. I. The Canonical Theory." Phys. Rev. {\bf 160} 1113 (1967).

\bibitem{G68}                 M. Gerstenhaber, ``On the Deformation of Rings and Algebras. III.", Ann. Math. {\bf 88} 1 (1968).  

%
\bibitem{Yano70}              K. Yano, {\it Integral Formulas in Riemannian Geometry} (Dekker, New York 1970).	
	
\bibitem{Klauder70}           J.R. Klauder, ``Ultralocal Scalar Field Models", Commun. Math. Phys. {\bf 18} 307 (1970). 

\bibitem{York73}              J.W. York Jr., ``Conformally Invariant Orthogonal Decomposition of Symmetric Tensors on Riemannian Manifolds and the 
                              Initial-Value Problem of General Relativity", J. Math. Phys. {\bf 14} 456 (1973).

\bibitem{T73}                 C. Teitelboim, ``How Commutators of Constraints Reflect Spacetime Structure", Ann. Phys. N.Y. {\bf 79} 542 (1973).   
	
												 
\bibitem{I76}                 C.J. Isham, ``Some Quantum Field Theory Aspects of the Superspace Quantization of General Relativity", Proc. R. Soc. Lond. {\bf A351} 209 (1976). 
              
\bibitem{L78}                 F. Bayen, M. Flato, C. Fronsdal, A. Lichnerowicz and D. Sternheimer, 
                              {\it Deformation Theory and Quantization. I. Deformations of Symplectic Structures} Ann. Phys. {\bf 111} 61 (1978).
							  
\bibitem{Henneaux79}          M. Henneaux, ``Geometry of Zero Signature Space-Times", Bull. Soc. Math. Belg. {\bf 31} 47 (1979). 
															  							
%
\bibitem{Teitelboim}          C. Teitelboim, ``The Hamiltonian Structure of Spacetime", 
                              in {\it General Relativity and Gravitation: One Hundred Years after the Birth of Albert Einstein} Vol 1 ed. A. Held (Plenum Press, New York 1980).

\bibitem{John}                F. John, {\it Partial Differential Equations} (Springer, New York 1982). 

\bibitem{BT82}                R. Bott and L. Tu, {\it Differential Forms in Algebraic Topology} (Springer, New York 1982).
		
\bibitem{I84}                 C.J. Isham, ``Topological and Global Aspects of Quantum Theory", 
                              in {\it Relativity, Groups and Topology {II}}, ed. B. DeWitt and R. Stora (North-Holland, Amsterdam 1984).

\bibitem{I89-I91}             C.J. Isham, ``Quantum Topology and Quantization on the Lattice of Topologies", Class. Quan. Grav {\bf 6} 1509 (1989);  

                              ``Quantization on the Lattice of Topologies, in {\it Florence 1989, Proceedings, Knots, Topology and Quantum Field Theories} 
							   ed. L. Lusanna (World Scientific, Singapore 1989);    

                              ``An Introduction To General Topology And Quantum Topology, unpublished, Lectures given at Banff in 1989 
							   (available on the KEK archive); 

							  C.J. Isham, Y.A. Kubyshin and P. Renteln, ``Quantum Norm Theory and the Quantization of Metric Topology", 
							  Class. Quant. Grav. {\bf 7} 1053 (1990);   					  
								
							 ``Quantum Metric Topology", in {\it Moscow 1990, Proceedings, Quantum Gravity} 
                              ed M.A. Markov, V.A. Berezin and V.P. Frolov (World Scientific, Singapore 1991); 
								
							  C.J. Isham, ``Canonical Groups And The Quantization Of Geometry And Topology", in {\it Conceptual Problems of Quantum Gravity} ed. 
                              A. Ashtekar and J. Stachel (Birkh\"{a}user, Boston, 1991).  
							  
%
\bibitem{HTBook}              M. Henneaux and C. Teitelboim, {\it Quantization of Gauge Systems} (Princeton University Press, Princeton 1992).   

\bibitem{K92}                 K.V. Kucha\v{r}, ``Time and Interpretations of Quantum Gravity", 
                              in {\it Proceedings of the 4th Canadian Conference on General Relativity and Relativistic Astrophysics} 
                              ed. G. Kunstatter, D. Vincent and J. Williams (World Scientific, Singapore 1992).  
							  							  
\bibitem{I93}                 C.J. Isham, ``Canonical Quantum Gravity and the Problem of Time",
                              in {\it Integrable Systems, Quantum Groups and Quantum Field Theories}  
                              ed. L.A. Ibort and M.A. Rodr\'{\i}guez (Kluwer, Dordrecht 1993), gr-qc/9210011.

\bibitem{K93}                 K.V. Kucha\v{r}, ``Canonical Quantum Gravity", in {\it General Relativity and Gravitation 1992},  
                              ed. R.J. Gleiser, C.N. Kozamah and O.M. Moreschi M (Institute of Physics Publishing, Bristol 1993), gr-qc/9304012.
							  							  
\bibitem{Olver2}              P.J. Olver, {\it Equivalence, Invariants and Symmetry} (C.U.P. 1995).
						
\bibitem{Bertlmann}           R.A. Bertlmann, {\it Anomalies in Quantum Field Theory} (Clarendon, Oxford 1996).
						
\bibitem{MT}                  I. Madsen and J. Tornehave, {\it From Calculus to Cohomology: De Rham Cohomology and Characteristic Classes} (C.U.P. 1997).  

\bibitem{Kontsevich}          M. Kontsevich,  ``Deformation Quantization of Poisson Manifolds, I.", Lett. Math. Phys. {\bf 66} 157 (2003), q-alg/9709040.  
							 
\bibitem{S98}                 D. Sternheimer, ``Deformation Quantization: Twenty Years After (AIP Conference Proceedings, 1998).  
							 							 
%
\bibitem{RWR}                 J.B. Barbour, B.Z. Foster and N. \'{o} Murchadha, ``Relativity Without Relativity", 
                              Class. Quant. Grav. {\bf 19} 3217 (2002), gr-qc/0012089.

\bibitem{Hatcher}             A. Hatcher, {\it Algebraic Topology} (Cambridge University Press, Cambridge 2001).

\bibitem{GB01}                M. Goze, J.M.A. Bermudez, ``On the Classification of Rigid Lie Algebras", J. Algebra {\bf 245} 68 (2001). 

\bibitem{OM02}                N. \'{o} Murchadha,  ``Constrained Hamiltonians and Local-Square-Root Actions", Int. J. Mod. Phys {\bf A20} 2717 (2002). 
								  
\bibitem{San}                 E. Anderson, ``Strong-Coupled Relativity without Relativity", Gen. Rel. Grav. {\bf 36} 255 (2004), gr-qc/0205118. 

\bibitem{I-Cat}               C.J. Isham, ``A New Approach to Quantising Space-Time: I. Quantising on a General Category", Adv. Theor. Math. Phys. {\bf 7} 331, gr-qc/0303060; 

                                          ``A New Approach to Quantising Space-Time: II. Quantising on a Category of Sets" {\bf 7} 807 (2003), gr-qc/0304077; 
	
	                                      ``A New Approach to Quantising Space-Time: III. State Vectors as Functions on Arrows" {\bf 8} 797 (2004), gr-qc/0306064; 
 						
	 						               ``Quantising on a Category", quant-ph/0401175.  

\bibitem{CM}                  M. Crainic and I. Moerdijk, ``Deformations of Lie Brackets: Cohomological Aspects", 
                              J. European Math. Soc. {\bf 10} 4 (2008), arXiv:math/0403434.  
							  										  
\bibitem{Lan}                 E. Anderson, ``Leibniz--Mach Foundations for GR and Fundamental Physics", 
                              in {\it Progress in General Relativity and Quantum Cosmology Research} ed A. Reimer (Nova, New York), gr-qc/0405022.  

\bibitem{Landsman}            N.P. Landsman, ``Between Classical and Quantum", in {\it Handbook of the Philosophy of Physics} (Elsevier, 2005) quant-ph/0506082. 
							  
\bibitem{Phan}                E. Anderson, ``On the Recovery of Geometrodynamics from Two Different Sets of First Principles", 
                              Stud. Hist. Phil. Mod. Phys. {\bf 38} 15 (2007), gr-qc/0511070.
							  
\bibitem{Gilmore}             R. Gilmore, {\it Lie Groups, Lie Algebras, and Some of Their Applications} (Dover, New York 2006).  
 					
\bibitem{G06}                 M. Goze,  ``Lie Algebras : Classification, Deformations and Rigidity", 
                              Lessons given during the {\it Cinqui$\grave{\me}$me Ecole de G\'{e}om\'{e}trie Diff\'{e}rentielle et Syst$\grave{\me}$mes Dynamiques}, 
							  ENSET ORAN (Algeria), November 4-11, 2006, arXiv:math/0611793.   

\bibitem{M08}                 P.W. Michor, ``Topics in Differential Geometry" (A.M.S.,  2008). 
							  							  
\bibitem{Giu09}               D. Giulini, ``The Superspace of Geometrodynamics", Gen. Rel. Grav. {\bf 41} 785 (2009) 785, arXiv:0902.3923.  
	
%
\bibitem{APoT}                E. Anderson, in {\it Classical and Quantum Gravity: Theory, Analysis and Applications} 
                              ed. V.R. Frignanni (Nova, New York 2011), arXiv:1009.2157. 
							  
\bibitem{PVM}                 P. Vargas Moniz, {\it Quantum Cosmology -- The Supersymmetric Perspective -- Vols. 1 and 2} (Springer, Berlin 2010).  

\bibitem{BojoBook}           M. Bojowald, {\it Canonical Gravity and Applications: Cosmology, Black Holes, and Quantum Gravity} (Cambridge University Press, Cambridge 2011).   
 						  
\bibitem{APoT2}               E. Anderson, Annalen der Physik, {\bf 524} 757 (2012), arXiv:1206.2403.   

\bibitem{Bojo12}              M. Bojowald, ``A Loop Quantum Multiverse?", AIP Conf. Proc. {\bf 1514} 21 (2012), arXiv:1212.5150.	

\bibitem{Lee2}                J.M. Lee, {\it Introduction to Smooth Manifolds} 2nd Ed. (Springer, New York 2013).

\bibitem{AM13}                E. Anderson and F. Mercati, ``Classical Machian Resolution of the Spacetime Construction Problem", arXiv:1311.6541. 
 
\bibitem{Olver}               P.J. Olver, {\it Applications of Lie Groups to Differential Equations} 2nd Ed. (Springer, 2013).
  
\bibitem{AObs}                E. Anderson, ``Beables/Observables in Classical and Quantum Gravity", SIGMA {\bf 10} 092 (2014), arXiv:1312.6073. 
							  							  							  
\bibitem{APoT3}               E. Anderson, ``Problem of Time and Background Independence: the Individual Facets", arXiv:1409.4117.    
 		
\bibitem{ASoS}                E. Anderson, ``Spaces of Spaces", arXiv.1412.0239.
				
\bibitem{Bojo16}              M. Bojowald, S. Brahma, U. Buyukcam, F. D'Ambrosio ``Hypersurface-Deformation Algebroids and Effective Space-Time Models", 
                              Phys. Rev. D 94, 104032 (2016), arXiv:1610.08355.
								
\bibitem{ABook}               E. Anderson, {\it Problem of Time. Quantum Mechanics versus General Relativity}, (Springer International 2017) Found. Phys. {\bf 190};  
                              free access to its extensive Appendices is at https://link.springer.com/content/pdf/bbm
		
\bibitem{ALett}  	          E. Anderson, ``A Local Resolution of the Problem of Time", arXiv:1809.01908.     

\bibitem{A-Brackets}          E. Anderson,  ``Geometry from Brackets Consistency", arXiv:1811.00564.   
		
\bibitem{A-Killing}           E. Anderson, ``Shape Theories. I. Their Diversity is Killing-Based and thus Nongeneric", arXiv:1811.06516.  

\bibitem{A-Cpct}              E. Anderson, ``Shape Theories II. Compactness Selection Principles", arXiv:1811.06528. 
		
\bibitem{A-CBI}               E. Anderson, ``Shape Theory. III. Comparative Theory of Backgound Independence", arXiv:1812.08771. 

\bibitem{I}                   E. Anderson,  ``A Local Resolution of the Problem of Time. I. Introduction and Temporal Relationalism", arXiv:1905.06200.  

\bibitem{II}                  E. Anderson,  ``A Local Resolution of the Problem of Time. II. Configurational Relationalism", arXiv:1905.06206.  

\bibitem{III}                 E. Anderson,  ``A Local Resolution of the Problem of Time. III. The other aspects piecemeal", arXiv:1905.06212.  

\bibitem{IV}                  E. Anderson,  ``A Local Resolution of the Problem of Time. IV. Quantum outline and piecemeal Conclusion", arXiv:1905.06294.  

\bibitem{V}                   E. Anderson,  ``A Local Resolution of the Problem of Time. V. Combining Temporal and Configurational Relationalism for Finite Theories", 
                              arXiv:1906.03630.  
							  
\bibitem{VI}                  E. Anderson,  ``A Local Resolution of the Problem of Time. VI. Combining Temporal and Configurational Relationalism for Field Theories and GR", 
                              arXiv:1906.03635.  
							  
\bibitem{VII}                 E. Anderson,  ``A Local Resolution of the Problem of Time. VII. Constraint Closure", arXiv:1906.03641.

\bibitem{VIII}                E. Anderson,  ``A Local Resolution of the Problem of Time. VIII. Expression in Terms of Observables", forthcoming.

\bibitem{IX}                  E. Anderson,  ``A Local Resolution of the Problem of Time. IX. Spacetime Reconstruction", arXiv:1906.03642.

\bibitem{X}                   E. Anderson,  ``A Local Resolution of the Problem of Time. X. Spacetime Relationalism",  forthcoming.

\bibitem{XI}                  E. Anderson,  ``A Local Resolution of the Problem of Time. XI. Slightly Inhomogeneous Cosmology",  forthcoming.

\bibitem{XII}                 E. Anderson,  ``A Local Resolution of the Problem of Time. XII. Foliation Independence",  forthcoming.

\bibitem{XIII}                E. Anderson,  ``A Local Resolution of the Problem of Time. XIII. Classical combined aspects' Conclusion", forthcoming.

\bibitem{XIV}                 E. Anderson,  ``A Local Resolution of the Problem of Time. XIV. Supporting account of Lie's Mathematics", forthcoming.
		  							 
\end{thebibliography}
\end{document}